\documentclass[aps,prd,floatfix,nofootinbib,showpacs,twocolumn,superscriptaddress]{revtex4-1}

\usepackage{mathrsfs}
\usepackage{amsmath}
\usepackage{amssymb}
\usepackage{bbold}
\usepackage{color}
\usepackage{graphicx}
\usepackage{soul}

\usepackage[mathscr,scaled=1.15]{urwchancal}
\DeclareFontFamily{OT1}{pzc}{}
\DeclareFontShape{OT1}{pzc}{m}{it}%
{<-> s * [1.15] pzcmi7t}{}
\DeclareMathAlphabet{\mathpzc}{OT1}{pzc}{m}{it}

\definecolor{purple}{rgb}{0.5,0,0.5}
\definecolor{blue}{rgb}{0.0,0,0.9}
\usepackage[hypertex]{hyperref}
\hypersetup{colorlinks=true, linkcolor=purple, citecolor= purple, urlcolor=blue}

\begin{document}

\title{Distribution amplitudes of radially-excited \mbox{\boldmath$\pi$}- and \mbox{\boldmath$K$}-mesons}

\author{B.-L. Li}
\email{libolin0626@126.com}
\affiliation{Department of Physics, Nanjing University, Nanjing 210093, China}

\author{L. Chang}
\email{lei.chiong@gmail.com}
\affiliation{School of Physics, Nankai University, Tianjin 300071, China}

\author{F. Gao}
\email{hiei@pku.edu.cn}
\affiliation{Department of Physics and State Key Laboratory of Nuclear Physics and Technology, Peking University, Beijing 100871, China}
\affiliation{Collaborative Innovation Center of Quantum Matter, Beijing 100871, China}

\author{C.\,D.~Roberts}
\email{cdroberts@anl.gov}
\affiliation{Physics Division, Argonne National Laboratory, Argonne
Illinois 60439, USA}

\author{S.\,M. Schmidt}
\email{s.schmidt@fz-juelich.de}
\affiliation{Institute for Advanced Simulation, Forschungszentrum J\"ulich and JARA, D-52425 J\"ulich, Germany}

\author{H.-S. Zong}
\email{zonghs@nju.edu.cn}
\affiliation{Department of Physics, Nanjing University, Nanjing 210093, China}
\affiliation{State Key Laboratory of Theoretical Physics, Institute of Theoretical Physics, CAS, Beijing, 100190, China}

\date{2 June 2016}

\begin{abstract}
A symmetry-preserving truncation of the two-body bound-state problem in relativistic quantum field theory is used to compute the leading-twist parton distribution amplitudes (PDAs) for the first radial excitations of the $\pi$- and $K$-mesons.  In common with ground states in these channels, the PDAs are found to be dilated with respect to the relevant conformal-limit form and skewed toward the heavier valence-quark in asymmetric systems.  In addition, the PDAs of radially-excited pseudoscalar mesons are not positive definite, owing to the fact that dynamical chiral symmetry breaking (DCSB) forces the leptonic decay constant of such states to vanish in the chiral limit.  These results highlight that DCSB is expressed visibly in every pseudoscalar meson constituted from light-quarks.  Hence, so long as its impact is empirically evident in the pseudoscalar members of a given spectrum level, it is unlikely that chiral symmetry is restored in any of the hadrons that populate this level.
\end{abstract}



\maketitle

\section{Introduction}
One of the most challenging problems in contemporary physics is presented by the question: whence the mass of a hadron and hence that of the bulk of visible material in the Universe?  Numerical simulations of lattice-regularised quantum chromodynamics (QCD) have produced information on the hadron mass spectrum \cite{Dudek:2010wm, Mahbub:2012ri, Engel:2013ig, Alexandrou:2013fsu}; but such analyses do not readily supply an intuitive understanding of the origin of that mass and its distribution within hadrons.  Notwithstanding that, it is clear the answer does not lie with the Higgs boson, for if one measures its contribution to the proton mass, $m_p$, via the values it generates for the valence-quark current-masses, which explicitly violate both the conformal invariance and chiral symmetry of classical QCD, the Higgs is responsible for less-than 2\% of $m_p$.  Instead, dynamical chiral symmetry breaking (DCSB) is the key \cite{Nambu:2009zza}.

DCSB is a crucial emergent phenomenon in the Standard Model of Particle Physics.  It is quite probably tied closely to the confinement of gluons and quarks, and also simultaneously responsible for both the Nambu-Goldstone boson character of the (almost) massless pion and the roughly 1\,GeV value of $m_p$.  Uncovering the manner by which these features of Nature are realised has long been a subject of intense experimental and theoretical activity, some of which is reviewed in Refs.\,\cite{Roberts:2015lja, Horn:2016rip}.  Much of this work has focused on the ground-state pion, its structure and interactions.  For instance, it has revealed that DCSB is responsible for a marked broadening of this meson's leading-twist parton distribution amplitude (PDA) \cite{Chang:2013pq, Cloet:2013tta}, and also those of other meson ground-states \cite{Segovia:2013eca, Gao:2014bca, Shi:2015esa}, an effect which provides a plausible explanation of modern data on pion elastic and transition form factors \cite{Chang:2013nia, Brodsky:2011xx, Stefanis:2012yw, Raya:2015gvaPRD}.  The impact of DCSB on the properties of hadron excited states is less well explored and understood.

Of particular interest is the fact that, owing to DCSB, Nambu-Goldstone modes are the only pseudoscalar mesons to possess a nonzero leptonic decay constant in the chiral limit: the decay constants of their radial excitations vanish \cite{Dominguez:1976ut, Dominguez:1977nt, Kataev:1982xu, Volkov:1996br, Elias:1997ya, Volkov:1999yi, Andrianov:1998kj, Maltman:2001gc, Holl:2004fr, Holl:2005vu, Lucha:2006rq, McNeile:2006qy, Boyle:2013jp, Mastropas:2014fsa, Ballon-Bayona:2014oma, Jiang:2015paa}:
\begin{equation}
\label{fpin0}
\forall n \geq 1 \,, \; f_{\pi_n} \stackrel{\hat m = 0}{\equiv} 0 \,,
\end{equation}
where $n$ is the radial quantum number ($n=0$ is the ground state) and $\hat m$ is the renormalisation-group-invariant (RGI) current-quark mass.  This result follows from the general form of the Gell-Mann--Oakes-Renner relation for isospin-nonzero pseudoscalars \cite{GellMann:1968rz, Maris:1997hd, Qin:2014vya}:
\begin{equation}
\label{GMOR}
f_{M_5} \, m_{M_5}^2 = (\hat m^{M_5}_1 + \hat m^{M_5}_2) \hat \rho_{M_5}\,,
\end{equation}
where $M_5$ labels the meson, $\hat m^{M_5}_{1,2}$ are the current-masses of its valence-quarks, $f_{M_5}$ is the meson's leptonic decay constant, obtained from the pseudovector projection of its Bethe-Salpeter wave function onto the origin in configuration space, and $\hat \rho_{M_5}$ is the RGI analogue obtained via pseudoscalar projection \cite{Brodsky:2010xf, Chang:2011mu, Brodsky:2012ku, Chang:2013epa}.

Eqs.\,\eqref{fpin0}, \eqref{GMOR} must both be a natural outcome in any framework with a genuine connection to QCD.  If fine-tuning is required to achieve either of these features in a given approach, then that approach is inconsistent with basic dynamics, symmetries and symmetry-breaking patterns of QCD.  For example, it is straightforward to achieve $m_{\pi_0}=0$ in quantum mechanical models, but impossible to express the quadratic growth of $m_{\pi_0}$ with current-quark mass, \emph{i.e}. the complete content of Eq.\,\eqref{GMOR}.  Equally, models founded in quantum mechanics typically produce a suppression of the decay constants of radially excited states, owing to zeros in the associated bound-state wave functions, \emph{e.g}.\ a Schr\"odinger equation treatment of positronium $\gamma\gamma$-decays yields the following pattern of decay-strengths relative to the ground-state: $1/8$, $n=1$; $2/27$, $n=2$; etc.  However, such models cannot yield a vanishing value for even one decay constant, much less all of them.

It is known that DCSB places severe constraints on the wave function of the ground-state pion \cite{Maris:1997hd, Qin:2014vya}; but our last few observations emphasise, via Eq.\,\eqref{fpin0}, that it must also impose extraordinary constraints on the chiral-limit wave function for every excited-state pseudoscalar meson.  Of course, the nature of a wave function in quantum field theory depends on the approach adopted for its analysis; and only wave functions defined using light-front quantisation can strictly provide a connection between dynamical properties of the underlying relativistic quantum field theory and notions familiar from nonrelativistic quantum mechanics \cite{Keister:1991sb, Coester:1992cg, Brodsky:1997de}.  With a light-front wave function in hand, however, one can translate features that arise purely through the infinitely-many-body nature of relativistic quantum field theory into images whose interpretation is seemingly more straightforward.

A natural framework for deriving Eqs.\,\eqref{fpin0}, \eqref{GMOR}, and elucidating and expressing their impression on hadron structure and interactions, is provided by the symmetry-preserving analysis of QCD's Dyson-Schwinger equations (DSEs) \cite{Roberts:2015lja}.  This approach yields Poincar\'e-covariant Bethe-Salpeter wave functions, which do not have a probability interpretation.  However, methods have recently been developed which enable these covariant wave functions to be projected onto the light-front \cite{Chang:2013pq}, supplying predictions for the leading-twist PDAs of ground-state mesons which are practically indistinguishable from those obtained by analysing simulations of lattice-QCD (lQCD) \cite{Cloet:2013tta, Segovia:2013eca, Shi:2015esa, Horn:2016rip}.   Thus it is now possible to express the parton content of Eq.\,\eqref{fpin0} in a manner which can place valuable constraints on all approaches that may directly be connected with the light-front.  In this connection, we focus herein on computing the leading-twist PDAs of the first radial excitations of the $\pi$- and $K$-mesons.

Our manuscript is composed as follows.  Section\,\ref{DSEBSE} describes calculations of the Bethe-Salpeter wave functions for the radially-excited $\pi$- and $K$-mesons, and their masses and leptonic decay constants.  Section~\ref{PDAsec} begins with a brief synopsis of the behaviour to be expected of PDAs associated with meson radial excitations in the absence of DCSB, and then continues with a detailed explanation of both the method by which these PDAs can be computed from symmetry-preserving DSE solutions and the results obtained therewith.  A summary and perspective are presented in Sec.\,\ref{summary}.

\section{\mbox{\boldmath $\pi$}- and \mbox{\boldmath $K$}- meson radial excitations}
\label{DSEBSE}
\subsection{Bound-state equations}
In order to reach our goal, we must first compute the Bethe-Salpeter amplitudes associated with the radially-excited $\pi$- and $K$-mesons; and to achieve that, it is necessary to settle on a truncation of QCD's DSEs.  As explained elsewhere \cite{Binosi:2016rxzd}, Eqs.\,\eqref{fpin0}, \eqref{GMOR} are guaranteed in any symmetry preserving truncation.  For our immediate purposes, therefore, it is sufficient to use the simplest; namely, rainbow-ladder (RL) truncation,\footnote{Concerning ground-state PDAs, results obtained using RL truncation can be compared with those produced by the most sophisticated approximation currently available, the so-called DB kernels \cite{Chang:2011ei, Binosi:2014aea}: despite noticeable quantitative differences, they agree qualitatively in all respects \cite{Chang:2013pq, Chang:2013epa, Shi:2015esa}.} in which case the renormalised gap- and Bethe-Salpeter-equations are, respectively:
\begin{eqnarray}
\nonumber S(p)^{-1} &=& Z_2 \,(i\gamma\cdot p + m^{\rm bm})\\
& +& Z_2^2 \int^\Lambda_{d\ell}\!\! {\cal G}(\ell)
\ell^2 D_{\mu\nu}^{0}(\ell)
\frac{\lambda^a}{2}\gamma_\mu S(p-\ell) \frac{\lambda^a}{2}\gamma_\nu ,
\label{rainbowdse} \rule{1em}{0ex}\\
\nonumber
\Gamma_M(k;P) &=&  - Z_2^2\int_{dq}^\Lambda\!\!
{\cal G}((k-q)^2)\, (k-q)^2 \, D_{\mu\nu}^{0}(k-q)\\
&& \times
\frac{\lambda^a}{2}\gamma_\mu S(q_+)\Gamma_M(q;P) S(q_-) \frac{\lambda^a}{2}\gamma_\nu ,
\label{ladderBSE}
\end{eqnarray}
where: $\int_{d\ell}^\Lambda:=\int^\Lambda \!\! \mbox{\footnotesize $\frac{d^4 \ell}{(2\pi)^4}$}$ represents a Poincar\'e-invariant regularisation of the integral, with $\Lambda$ the ultraviolet regularization mass-scale; $Z_2(\zeta,\Lambda)$ is the quark wave function renormalisation constant, with $\zeta$ the renormalisation scale; $D^{0}_{\mu\nu}(\ell)$ is the Landau-gauge free-gauge-boson propagator;\footnote{Landau gauge is used for many reasons \protect\cite{Bashir:2009fv}, for example, it is: a fixed point of the renormalisation group; that gauge for which sensitivity to model-dependent differences between \emph{Ans\"atze} for the fermion--gauge-boson vertex are least noticeable; and a covariant gauge, which is readily implemented in numerical simulations of lattice regularised QCD.}
one can choose $q_\pm=q\pm P/2$ without loss of generality in this Poincar\'e covariant approach; and
\begin{equation}
\label{GIR}
\ell^2 {\cal G}(\ell^2)
= \ell^2 {\cal G}_{\rm IR}(\ell^2) + 4\pi \tilde\alpha_{\rm pQCD}(\ell^2)
\end{equation}
specifies the interaction, with $\tilde\alpha_{\rm pQCD}(k^2)$ a bounded, monotonically-decreasing regular continuation of the perturbative-QCD running coupling to all values of spacelike-$\ell^2$, and ${\cal G}_{\rm IR}(\ell^2)$ an \emph{Ansatz} for the interaction at infrared momenta, such that ${\cal G}_{\rm IR}(\ell^2)\ll \tilde\alpha_{\rm pQCD}(\ell^2)$ $\forall \ell^2\gtrsim 2\,$GeV$^2$.  The nature of ${\cal G}_{\rm IR}(\ell^2)$ determines whether confinement and/or DCSB are realised in solutions of the gap equation, with the former expressed in the sense described, \emph{e.g}. in Sec.\,3 of Ref.\,\cite{Horn:2016rip}.

The gap equation yields a dressed-quark propagator, which has the general form:
\begin{equation}
 S(p) 
= Z(p^2,\zeta^2)/[ i\gamma\cdot p + M(p^2)] \,,
\label{SgeneralN}
\end{equation}
and can be obtained from Eq.\,(\ref{rainbowdse}) augmented by a renormalisation condition.  A mass-independent scheme is useful and can be implemented by fixing all renormalisation constants in the chiral limit.  Notably, the mass function, $M(p^2)$, is independent of the renormalisation point; and the renormalised current-quark mass is given by
\begin{equation}
\label{mzeta}
m^\zeta = Z_m(\zeta,\Lambda) \, m^{\rm bm}(\Lambda) = Z_4^{-1} Z_2\, m^{\rm bm},
\end{equation}
wherein $Z_4$ is the renormalisation constant associated with the Lagrangian's mass-term. Like the running coupling constant, this running mass is a familiar concept.  The RGI current-quark mass may be inferred via
\begin{equation}
\hat m = \lim_{p^2\to\infty} \left(\tfrac{1}{2}\ln [p^2/\Lambda^2_{\rm QCD}]\right)^{\gamma_m} M(p^2)\,,
\end{equation}
where $\gamma_m = 12/(33-2 N_f)$: $N_f$ is the number of quark flavours employed in computing the running coupling; and $\Lambda_{\rm QCD}$ is QCD's dynamically-generated RGI mass-scale.  The chiral limit is expressed by
\begin{equation}
\label{chirallimit}
\hat m = 0\,.
\end{equation}

The Bethe-Salpeter equation (BSE) is an eigenvalue problem for a meson's mass-squared, \emph{i.e}. in a given channel, Eq.\,(\ref{ladderBSE}) has solutions only at particular, isolated values of $P^2=-m_M^2$.  At these values, solving the equation produces the associated meson's Bethe-Salpeter amplitude.  Herein we consider isospin-nonzero pseudoscalar states,\footnote{Masses and other properties of charge-neutral pseudoscalar mesons are affected by the non-Abelian anomaly.  In the BSE context, this is discussed in Ref.\,\protect\cite{Bhagwat:2007ha}.} 
so that only the following amplitude is relevant:
\begin{align}
\label{GammaP}
&\Gamma_{M_5}(k;P) = \sum_{i=1}^4 \gamma_5 \tau_{0^-}^i(k,P)\, F_{M_5}^i(k;P), \\
&\tau_{0^-}^1 =  i \mbox{\boldmath $I$}_D,\;
\tau_{0^-}^2 = \gamma\cdot P,\;
\tau_{0^-}^3 = \gamma\cdot k \,,\;
\tau_{0^-}^4  = \sigma_{\mu\nu} P_\mu k_\nu \,.\quad
\end{align}
The canonical normalisation condition \cite{Nakanishi:1969ph, LlewellynSmith:1969az} constrains the bound-state to produce a pole with unit residue in the quark-antiquark scattering matrix; and the Bethe-Salpeter wave function is
\begin{equation}
\chi_{M_5}(k;P) = S^f(k_+) \, \Gamma_{M_5}(k;P)  \, S^g(k_-)\,,
\end{equation}
where $f$, $g$ describe, respectively, the meson's valence-quark and -antiquark.

To proceed, it remains only to specify the interaction, Eq.\,\eqref{GIR}.  We use that proposed in Ref.\,\cite{Qin:2011dd, Qin:2011xq}, \emph{viz}.,
\begin{equation}
\label{CalGQC}
{\cal G}(s) = \frac{8 \pi^2}{\omega^4} D \, {\rm e}^{-s/\omega^2}
+ \frac{8 \pi^2 \gamma_m\,{\cal F}(s)}{\ln [ \tau + (1+s/\Lambda_{\rm QCD}^2)^2]} ,
\end{equation}
where: $N_f=4$ in $\gamma_m$, $\Lambda_{\rm QCD}=0.234\,$GeV; $\tau={\rm e}^2-1$; and ${\cal F}(s) = \{1 - \exp(-s/[4 m_t^2])\}/s$, $m_t=0.5\,$GeV.  This interaction preserves the one-loop renormalisation-group behavior of QCD in the gap- and Bethe-Salpeter-equations \cite{Maris:1997tm}, it is consistent with modern DSE and lattice studies \cite{Boucaud:2011ug, Aguilar:2015bud}, and the infrared structure serves to ensure confinement and DCSB.  Notably, as illustrated in Refs.\,\cite{Qin:2011dd, Qin:2011xq}, the parameters $D$ and $\omega$ are not independent: with $D\omega=\,$constant, one can expect computed observables to be practically insensitive to $\omega$ on the domain $\omega\in[0.4,0.6]\,$GeV.  We use $\omega=0.5\,$GeV.

\subsection{Amplitudes, masses and decay constants}
A detailed analysis of ground and radially-excited isospin-one pseudoscalar mesons is presented in Ref.\,\cite{Qin:2011xq}.  We follow that study and use $D\omega = (1.1\,\mbox{GeV})^3$ in our analysis of the first radial excitations of the $\pi$- and $K$-mesons; and, as elsewhere \cite{Chang:2013pq, Cloet:2013tta, Chang:2013epa, Segovia:2013eca, Shi:2015esa}, we work with a renormalisation scale $\zeta=\zeta_2 := 2\,$GeV.  The radially-excited $K$-meson is constituted from valence quarks with RGI current-masses \cite{Shi:2015esa}
\begin{equation}
\label{currentmass}
\hat m_u=\hat m_d = 6.8\,\mbox{MeV}\,,\;
\hat m_s = 162\,\mbox{MeV}\,,
\end{equation}
which correspond to one-loop evolved values:
\begin{equation}
m_{u=d}^{\zeta_2}=4.7\,\mbox{MeV}, \; m_{s}^{\zeta_2}=112\,\mbox{MeV}.
\end{equation}
So as to fully illustrate the implications of Eq.\,\eqref{fpin0}, we employ the chiral limit, Eq.\,\eqref{chirallimit}, for our analysis of the radially-excited $\pi$-meson.

In order to obtain the mass and amplitude associated with the first radial excitation of the $\pi$- and $K$-mesons from Eqs.\,\eqref{rainbowdse}, \eqref{ladderBSE}, we employ the methods of Refs.\,\cite{Krassnigg:2003wy, Krassnigg:2009gd} to solve the equations and isolate the excited states. This procedure yields
\begin{subequations}
\label{excitedresults}
\begin{align}
m^0_{\pi_1} & = 1.26\,{\rm GeV},\; f^0_{\pi_1} = 0 \,, \\
m_{K_1} & = 1.39\,{\rm GeV},\; f_{K_1} = 6.7\,{\rm MeV},
\end{align}
\end{subequations}
where we have included a superscript ``0'' to emphasise that the $\pi_1$ results were obtained in the chiral limit.\footnote{At a realistic value of the current-quark mass,  Eq.\,\eqref{currentmass}, $f_{\pi_1} =1.6\,$MeV \cite{Holl:2004fr, Holl:2005vu}.   \emph{N.B}.\ It is fair to consider that RL-truncation used in connection with one-loop QCD renormalisation-group-improved kernels for the gap- and bound-state-equations is accurate at the level of 15\% \cite{Roberts:2007jh}.
}
Nonnegative values of $f_{\pi_1,K_1}$ result from our decision to employ a convention that produces a negative value at $k^2=0$ for the $j=0$ Chebyshev-moment of the $F^1$-term in the excited-state Bethe-Salpeter amplitudes:
\begin{equation}
\label{Chebyshev}
^j \!F_{M_5}(k^2) := \frac{2}{\pi} \int_{-1}^{1}\!\! dx\,\sqrt{1-x^2}\,U_j(x)\,F_{M_5}(k^2,x;P^2)\,,
\end{equation}
where $k\cdot P = x \sqrt{k^2 P^2}$ and $U_n(x)$ is a Chebyshev polynomial of the second kind.  When solving for the Bethe-Salpeter amplitude, we adopted the Chebyshev expansion technique described as ``Method B'' in Ref.\,\cite{Maris:1997tm}; and $f_{M_5}>0$ is guaranteed by normalising such that $^0 \!F_{M_5}(k^2\to \infty) \to 0^+$.

A context for the results in Eqs.\,\eqref{excitedresults} is provided by the following empirical values  \cite{Agashe:2014kda, Diehl:2001xe}:
\begin{align}
m_{\pi_1} & = 1.3(1)\,{\rm GeV},\; f_{\pi_1}< 5.9\,{\rm MeV}\,,\\
m_{K_1} & = 1.43(5)\,{\rm GeV}\,;
\end{align}
and $f_{\pi_1}=1.6(3)\,$MeV, $f_{K_1}=15(2)\,$MeV estimated using finite-energy sum-rules \cite{Maltman:2001gc}.  (We employ a normalisation with which the empirical values of the ground-state pion and kaon leptonic decay constants are $f^E_\pi=92\,$MeV, $f^E_K=110\,$MeV.)  It is notable, too, that the same framework predicts $f_{\rho_1}/f_{\rho_0} \approx 0.6$ \cite{Qin:2011xq}, whilst a sum-rules analysis yields $f_{\rho_1}/f_{\rho_0}=0.77(9)$ \cite{Jiang:2015paa}.

\begin{figure}[t]
\centerline{\includegraphics[width=0.45\textwidth]{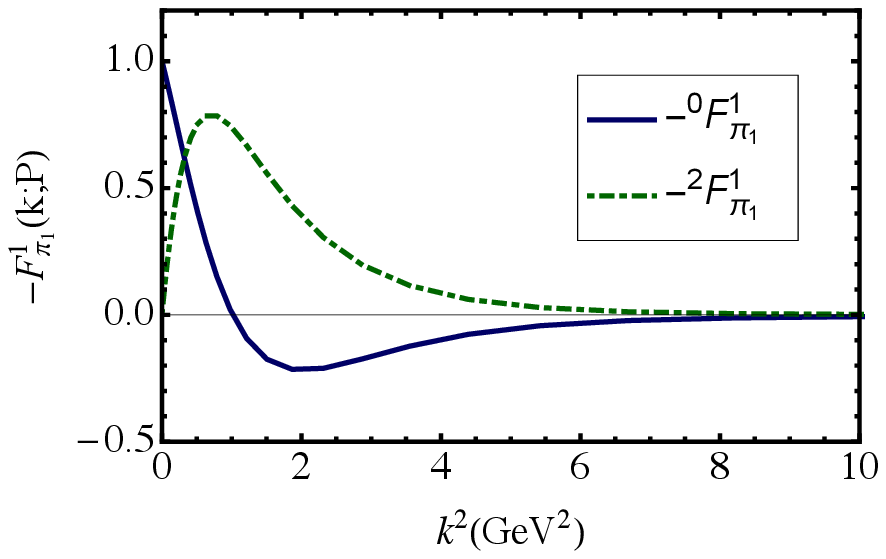}}

\centerline{\includegraphics[width=0.45\textwidth]{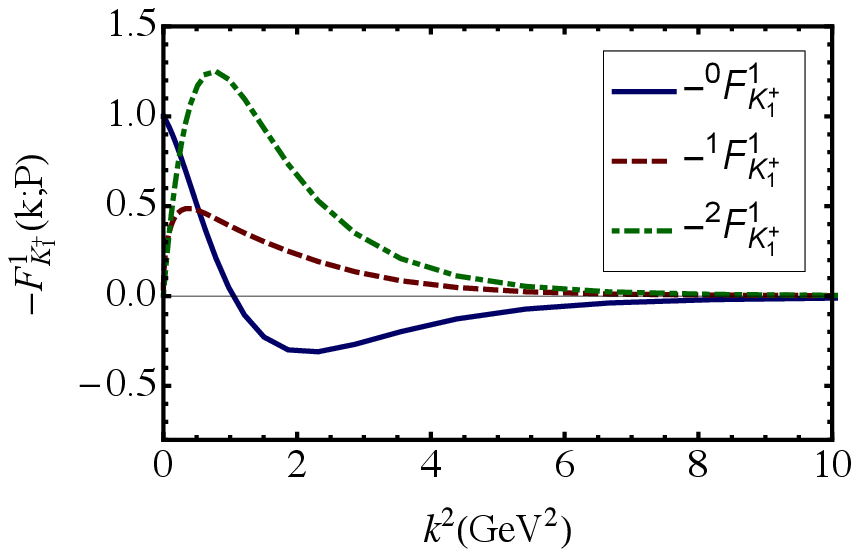}}
\caption{\label{BSamps0}
Leading Chebyshev moments obtained from the dominant piece of the $\pi_1$ and $K_1$ Bethe-Salpeter amplitudes, \emph{i.e}.\ $F_{M_5}^1$, $M_5=\pi_1^+$, $K_1^+$, in Eq.\,\eqref{GammaP}.  $j$-odd Chebyshev moments are zero for the $\pi_1$ because it is a charge-conjugation eigenstate.  Notably, in contrast to ground-states, the first few Chebyshev moments in radial excitations are all of comparable magnitude.  (\emph{N.B}.\ The figures display negative-$F^1$.)
}
\end{figure}

The leading Chebyshev-moments of the $F_{M_5}^{i=1}$ amplitudes associated with the $\pi$ and $K$ radial excitations are depicted in Fig.\,\ref{BSamps0}.  The appearance of a single zero in the $j=0$ Chebyshev moment is a characteristic feature of the amplitude associated with a meson's first radial excitation \cite{Holl:2004fr}.  It is particularly important to highlight here that the relative weighting of the domains of positive and negative support in the multi-component Bethe-Salpeter amplitude associated with the chiral-limit $\pi_1$-meson is precisely determined in a symmetry-preserving DSE solution such that, independent of any and all parameters,
\begin{equation}
\label{fpigen0}
f^0_{\pi_1} P_\mu  =
Z_2\; {\rm tr}_{\rm CD}
\int_{d\ell}^\Lambda i\gamma_5\gamma_\mu  \chi_{\pi_1}^0(\ell;P)   = 0\,,
\end{equation}
where $P$ is the meson's four-momentum, $P^2=-m_{\pi_1}^2$, and the trace is over colour and spinor indices.

\section{PDAS of the radial excitations}
\label{PDAsec}
\subsection{Expectations absent DCSB}
\label{AdSQCD}
We noted in the Introduction that quantum mechanical models typically produce a suppression of the leptonic decay constant for meson radial excitations because they introduce zeros in the excited-state wave functions; but they cannot make the decay constants vanish.  This is apparent, \emph{e.g}.\ in Ref.\,\cite{Arndt:1999wx}, which computes $f_{\pi_1}/f_{\pi_0} = 0.20$, $f_{\pi_2}/f_{\pi_0} = 0.46$.

Another example is provided by the holographic-QCD framework reviewed in Ref.\,\cite{Brodsky:2014yha}.  A model of mesons is described therein, along with, \emph{inter alia}, the associated light-front wave functions for all meson excitations, so it is straightforward to calculate the associated PDAs.  Consider, therefore, the model's chiral-limit wave functions for ground-state and radially-excited mesons:
\begin{subequations}
\label{AdSbspace}
\begin{align}
\nonumber
 \Psi_{\pi_n}(x,|b_\perp|)  &
=2 \kappa {\mathpzc X}(x)  \\
&  \times {\rm e}^{- b^2_\perp \kappa^2  {\mathpzc X}^2(x)/2} \, L_{n}(b_\perp^2 \kappa^2 {\mathpzc X}^2(x))\,,\\
{\mathpzc X}(x) & = \sqrt{x(1-x)}\,,
\end{align}
\end{subequations}
where $\kappa\approx 0.5\,$GeV is the model's mass-scale and $L_{n}$ is a Laguerre polynomial.  Notably, within this holographic model, these wave functions simultaneously represent the structure of $\pi_n$- and $\rho_n$-mesons $\forall n\geq 0$.

Eq.\,\eqref{AdSbspace} defines wave functions in impact-parameter space.  The momentum-space results are obtained via Fourier transform, \emph{e.g}.
\begin{subequations}
\begin{align}
\Psi_{\pi_0}(x,|k_\perp|) & =\tfrac{4\pi}{\kappa{\mathpzc X}(x)}\,
{\rm e}^{-k_\perp^2/[2\kappa^2{\mathpzc X}^2(x)]}\,,\\
\Psi_{\pi_1}(x,|k_\perp|) & =
\Psi_{\pi_0}(x,|k_\perp|)\,\frac{(k_\perp^2-\kappa^2{\mathpzc X}^2(x))}{\kappa^2 {\mathpzc X}^2(x)}\,,
\end{align}
\end{subequations}
with expressions of greater complexity for $n\geq 2$.  Owing to the presence of a single zero and since $\mathpzc{X}\geq 0$, $\Psi_{\pi_1}(x,|k_\perp|=0)\leq 0$, a result consistent with the convention we adopted for Eq.\,\eqref{excitedresults}.  Moreover, orthonormality  is here guaranteed via the $k_\perp$-integral alone:
\begin{equation}
\label{kporthogonal}
\int \frac{d^2 k_\perp}{16\pi^3} \, \Psi_{\pi_i}(x,|k_\perp|)\, \Psi_{\pi_j}(x,|k_\perp|) = \delta_{ij}\,.
\end{equation}

The PDA associated with a given wave function in Eq.\,\eqref{AdSbspace} is:
\begin{subequations}
\label{phiAQ}
\begin{align}
\phi_{\pi_n}^{hQ}(x) & = \int \frac{d^2 k_\perp}{16 \pi^3}
\int d^2 b_\perp{\rm e}^{i k_\perp \cdot b_\perp} \, \Psi_{\pi_n}(x,|b_\perp|)  \\
& = \Psi_{\pi_n}(x,|b_\perp|=0) = \tfrac{\kappa}{2\pi}\,{\mathpzc X}(x)\,,
\label{nforall}
\end{align}
\end{subequations}
\emph{viz}.\ the same result $\forall n\geq 0$.  Stated plainly: in the holographic model of Ref.\,\cite{Brodsky:2014yha}, the PDA of every one of the pion's radial excitations is \emph{identical}.  

At this point we would like to reiterate that, in the chiral limit, the light-front holographic model predicts the same wave functions for all $\pi$- and $\rho$-mesons.  In constructing $\rho$-meson PDAs, however, one must amend Eq.\,\eqref{phiAQ} by including information about the quark and antiquark helicites in this $J=1$ system \cite{Dosch:1996ss, Ahmady:2012dy}.  Doing that, one may arrive at ground-state $\rho$-meson PDAs that are broadly consistent with other analyses, \emph{e.g}.\ obtaining two independent leading-twist PDAs associated with $\rho$-mesons \cite{Ball:1996tb}: $\phi_\rho^\|(x)$ and $\phi_\rho^\perp(x)$,  which are connected, respectively, with a description of the light-front fraction of the $\rho$-meson's total momentum carried by the quark in a longitudinally or transversely polarised bound-state.  Notably, QCD-connected calculations indicate that $\phi_{\rho_0}^\|(x)$ is significantly narrower than $\phi_{\rho_0}^\perp(x)$, which itself is much narrower than $\phi_{\pi_0}(x)$ \cite{Pimikov:2013usa, Gao:2014bca}.

With the conventions employed in Ref.\,\cite{Brodsky:2014yha} and using Eq.\,\eqref{nforall}, one finds the following results for the leptonic decay constants of pseudoscalar mesons:
\begin{equation}
\forall n \geq 0\,, \; f_{\pi_n} = 2 \sqrt{N_c} \int_0^1dx \, \phi^{hQ}_{\pi_0}(x) \,,
\end{equation}
and hence the holographic model predicts ($N_c=3$)
\begin{equation}
\forall n \geq 0\,, \;
f_{\pi_n} = f_{\pi_0}  =  \frac{\surd 3}{8} \kappa \sim 0.11\,{\rm GeV}\,,
\label{holofpin}
\end{equation}
an outcome in marked conflict with Eq.\,\eqref{fpin0} $\forall n\geq 1$.%
\footnote{It has been argued \cite{Brodsky:2011xx} that for $n=0$ one should include a factor $Z^{q \bar q}\approx 0.94$ in Eq.\,\eqref{holofpin} so as to express the probability of finding the valence $q\bar q$-component in the physical pion at the model's scale.  The remainder then reflects the presence of higher Fock-space components, \emph{viz}.\ a ``meson cloud''.  The value of $Z^{q \bar q}$ was not calculated, but, instead, determined through a fit to pion electromagnetic form factor data (Ref.\,\cite{Brodsky:2014yha}, Sec.\,6.1.5).   It is conceivable that a similar factor, $Z^{q \bar q}\to Z_n^{q \bar q}$, should appear for every one of the holographic model's radial excitations; and this could alter the conclusion in Eq.\,\eqref{holofpin} \cite{GFdTPrivate}.  However, it is unlikely to mend the basic conflict between Eq.\,\eqref{fpin0} and the formulation of the holographic model in Ref.\,\cite{Brodsky:2014yha}.  Only a very particular symmetry-breaking pattern can enforce $Z_n^{q \bar q} = 0 $ $\forall n\geq 1$, and consistency between the holographic formulation in Ref.\,\cite{Brodsky:2014yha} and the axial-vector Ward-Green-Takahashi identity has yet to be demonstrated.  On the other hand, that has been achieved in a different holographic model \cite{Ballon-Bayona:2014oma}.}

The result in Eq.\,\eqref{phiAQ} is supposed to be valid at a scale appropriate to the AdS/QCD model, which is typically assumed to be $\zeta_{hQ} \sim 1\,$GeV.  Given that the ERBL evolution equations \cite{Lepage:1979zb, Efremov:1979qk, Lepage:1980fj} for pseudoscalar-meson radial excitations are the same as those for ground-states,  then the conformal limit result for these PDAs can be written:
\begin{equation}
\label{phicl}
\forall n\geq 0\,,\;
\phi_{\pi_n}^{hQ}(x) \stackrel{\Lambda_{\rm QCD}/\zeta \simeq 0}{\approx} \hat\phi^{\rm cl}(x) = f_{\pi_0}\, \surd{3} \, x (1-x)\,.
\end{equation}

Naturally, in a truly conformal theory, all mass-scales disappear: there is no dynamics in a conformal theory, only kinematics, and hence bound-states are impossible.  In connection with the material in this subsection, that is evident via $\kappa\to 0 \Rightarrow  f_{\pi_0}\to 0$, so that all the PDAs and, indeed, all the light-front wave functions vanish.  

\subsection{DSE results}
With symmetry-consistent solutions of the gap- and Bethe-Salpeter-equations in hand, the leading-twist PDA of any given pseudoscalar meson, $\phi_{M_5}(x)$, can be obtained via the following light-front projection:
\begin{equation}
 \phi_{M_5}(x) = {\rm tr}_{\rm CD}
Z_2 \! \int_{d\ell}^\Lambda \!\!
\delta_n^x(\ell_+)\,\gamma_5\gamma\cdot n\, \chi_{M_5}(\ell;P)\,,
\label{pionPDA}
\end{equation}
where $\delta_n^x(\ell_+)=\delta(n\cdot \ell_+ - x n\cdot P)$, $n^2=0$, $n\cdot P = -m_{M_5}$, and $\phi_{M_5}$ has mass-dimension one in this convention, as in Sec.\,\ref{AdSQCD}; but, using our canonical normalisation,
\begin{equation}
\label{CDRcanonical}
\phi^{\rm cl}(x) = f_{\pi_0} \, 6 x(1-x)\,.
\end{equation}
The Bethe-Salpeter wave function is typically computed in Euclidean space, whereafter one can reconstruct the PDA from its Mellin moments \cite{Chang:2013pq, Gao:2014bca, Shi:2015esa, Ding:2015rkn}:
\begin{equation}
\langle x^m\rangle   := \int_0^1 dx \, x^m \phi_{M_5}(x)\,,
\end{equation}
which are here given explicitly by
\begin{equation}
\langle x^m\rangle  =
{\rm tr}_{\rm CD} Z_2 \! \int_{d\ell}^\Lambda \!\! {\cal D}(n,\ell,P,m)
 \,\gamma_5\gamma\cdot n\,\chi_{M_5}(\ell;P)\,,
\label{phimom}
\end{equation}
where ${\cal D}(n,\ell,P,m) = (n\cdot\ell_+)^m/(n\cdot P)^{m+1}$.

\begin{figure}[t]
\centerline{\includegraphics[width=0.47\textwidth]{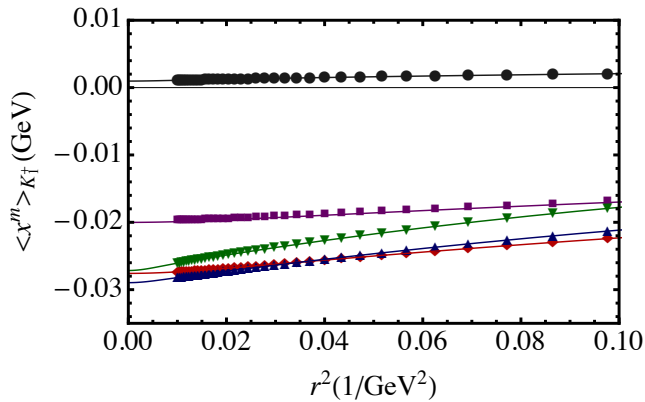}}
\caption{\label{extrap}
The $K_1^+$ moments in Table~\ref{Table:moments} ($m\geq 1$) are the $r^2\to 0$ extrapolations of the curves depicted in this figure:
$\langle x^1\rangle$, black circles;
$\langle x^2\rangle$, purple squares;
$\langle x^3\rangle$, red diamonds;
$\langle x^4\rangle$, blue-up-triangles;
$\langle x^5\rangle$, green down-triangles.
The curves are $[2,1]$-Pad\'e fits to the points depicted.  Other fitting forms were also employed, with no quantitative change in the results.  Naturally, no extrapolation is required for the $m=0$ moment, which is simply the meson's leptonic decay constant.
}
\end{figure}

When the meson's mass is small, \emph{viz}.\ $m_{M_5} \lesssim m_p$, the $(n\cdot \ell_+)^m$ factor in Eq.\,\eqref{phimom} produces a highly-oscillatory integrand and thus reliable values for the moments cannot be obtained using a direct approach to computing the integrals.  In these cases, the procedure of Ref.\,\cite{Chang:2013pq}, based upon generalised spectral representations of the light-quark propagators and bound-state amplitudes, is necessary and efficacious.  With increasing bound-state mass, however, owing to a damping influence from $n\cdot P$, this problem is shifted to progressively higher moments, which are also of diminishing magnitude and hence have little real impact.  Accordingly, a ``brute-force'' approach is feasible for radially excited states.

A practical implementation of the brute-force method is described in Ref.\cite{Ding:2015rkn} and we follow that technique; namely, direct computation of the integrals defined by Eq.\,\eqref{phimom} using interpolations of numerical solutions for the propagators and Bethe-Salpeter amplitudes.  In order to eliminate dependence on the upper-bound of the momentum integration, which is a remnant of the oscillation problem just described, a factor
\begin{equation}
\label{cutofffunction}
{\mathpzc d}(k^2 r^2) = 1/(1+k^2 r^2)^{m/2}
\end{equation}
is introduced for each $\langle x^{m}\rangle$, $m\geq 1$. The moment is then computed as a function of $r^2$, with the values subsequently fitted by a smooth function, which is used to extrapolate to $r^2=0$.  The reliability of this procedure is illustrated by Fig.\,\ref{extrap}; and the results are listed in Table~\ref{Table:moments}.  In all cases we found that reliable estimates could be obtained for $m\leq 5$.  Higher moments showed modest sensitivity to the number of Chebyshev moments retained in solving the BSE for the given radial excitation and were therefore discarded.  We verified that the same results are obtained using different forms of regulator function in Eq.\,\eqref{cutofffunction}.  The results listed in Table~\ref{Table:moments} were obtained with $j=0,\ldots,5$ in the Chebyshev expansion of each meson's Bethe-Salpeter amplitude, Eq.\,\eqref{Chebyshev}.

Using the moments in Table~\ref{Table:moments}, the PDAs of the pseudoscalar-meson radial-excitations, $\pi_1$, $K_1^+$, may be reconstructed using the method introduced and tested in Refs.\,\cite{Chang:2013pq, Cloet:2013tta, Chang:2013epa, Segovia:2013eca, Shi:2015esa}.  One first writes\footnote{A normalising constant, based on $a^0_{M_5}$, is not factorised in Eq.\,\eqref{Gfit} because $\langle x^0\rangle \equiv 0$ in the chiral limit.  Notwithstanding that, correct normalisation is guaranteed owing to the canonical procedure we have adopted for the Bethe-Salpeter amplitude.}
\begin{equation}
\label{Gfit}
\phi_{M_5}(x) = [x \bar x]^{\alpha_-} \sum_{z=0}^{z_{\rm max}} a^z_{M_5}C_z^\alpha(x-\bar x)\,,
\end{equation}
where $\{C_z^\alpha\}$ are order-$\alpha$ Gegenbauer polynomials, $\alpha_-=\alpha-1/2$, and $\bar x=(1-x)$.  (For a charge-conjugation eigenstate, like the $\pi_1$, the sum only includes even Gegenbauer polynomials.)
In this analysis, we use $z_{\rm max}=4$; and the parameters $\{\alpha,a_{M_5}^{z=0,1,\dots}\}$ are determined in a least-squares fit that requires the moments of $\phi_{M_5}(x) $ in Eq.\,\eqref{Gfit} to match those in Table~\ref{Table:moments}, with the results:
\begin{equation}
\label{PDFparams}
\begin{array}{l|cccccc}
           & \alpha & a^0 & a^1 & a^2 & a^3 & a^4 \\\hline
\pi_1   & 1.21    & 0     & 0     & 1.00 & 0    & 0.180 \\
K_1^+& 1.30    & 0.0285 & -0.0657 & 1.24 & -0.703 &  0.0813
\end{array}\,,
\end{equation}
where the coefficients are measured in GeV.  We have checked, and using moments computed from the
$K^-$ Bethe-Salpeter wave function we obtain
\begin{equation}
\phi_{K^-} (x) = \phi_{K^+}(\bar x)\,.
\end{equation}
Additionally, the results are practically unchanged if one changes $z_{\rm max}\to 6$.

\begin{table}[t]
\caption{Mellin moments $(\times 10^2)$ of the leading-twist $\pi_1^+$ and $K_1^+$ PDAs, computed from Eq.\,\eqref{phimom} using the method described in connection with Eq.\,\eqref{cutofffunction} and Fig.\,\ref{extrap}.  (The Bethe-Salpeter wave functions are computed in RL-truncation.)
\label{Table:moments}
}
\begin{tabular*}
{\hsize}
{
l|@{\extracolsep{0ptplus1fil}}
c|@{\extracolsep{0ptplus1fil}}
c|@{\extracolsep{0ptplus1fil}}
c|@{\extracolsep{0ptplus1fil}}
c|@{\extracolsep{0ptplus1fil}}
c|@{\extracolsep{0ptplus1fil}}
c@{\extracolsep{0ptplus1fil}}}\hline
$10^2 \langle x^m \rangle$ & $m=0$ & $1$ & $2$ & $3$ & $4$ & $5$ \\\hline
$\pi_1^+$ & $0$ & $0$ & 1.99 & 3.02 & 3.40 & 3.46\\
$K_1^+$ & $0.668$ & $-0.101$ & 2.00 & 2.76 & 2.86 & 2.66\\\hline
\end{tabular*}
\end{table}

\begin{figure}[t]
\centerline{\includegraphics[width=0.45\textwidth]{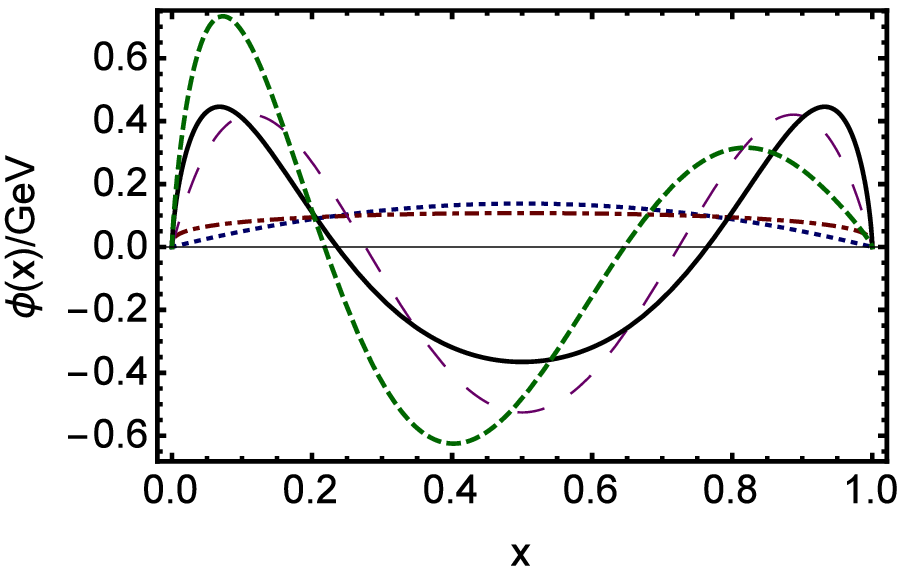}}
\centerline{\includegraphics[width=0.45\textwidth]{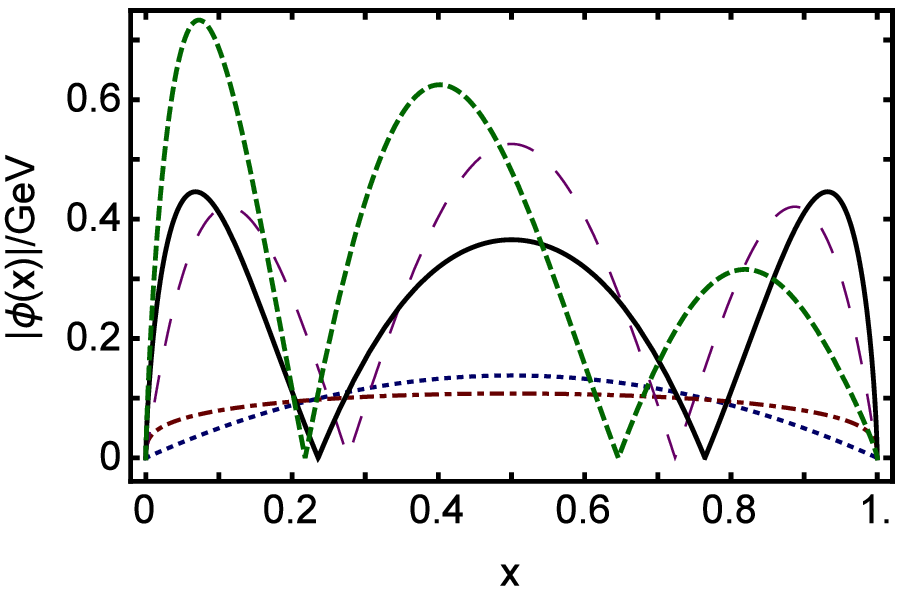}}
\caption{\label{PDAs} Leading-twist PDAs, first radial excitation of $\pi^+$- and $K^+$-mesons, computed at a resolving scale $\zeta_2=2\,$GeV: solid (black) curve, $\pi_1^+$; and dashed (green) curve, $K_1^+$.  For comparison: dotted (blue) curve, ground-state conformal limit result, Eq.\,\eqref{CDRcanonical}; dot-dashed (red) curve, RL-truncation result for ground-state $\pi$-meson, Eq.\,\eqref{phiRL}; and long-dashed (purple) curve, radial-excitation ``conformal limit'' function described in connection with Eq.\eqref{pincl}.  \emph{Upper panel} -- the PDA itself; and \emph{lower panel}, its absolute value.
}
\end{figure}

The PDAs defined by Eqs.\,\eqref{Gfit}, \eqref{PDFparams} are depicted in Fig.\,\ref{PDAs}.  For comparison, the PDA of the pion ground-state computed in RL-truncation is also drawn:
\begin{equation}
\label{phiRL}
\phi_{\pi_0}^{RL}(x;\zeta_2) \approx  f_{\pi_0}\,1.77 \, [x(1-x)]^{0.3}.
\end{equation}
It is worth reiterating that herein our convention is to define all PDAs such that they carry mass-dimension one.  This enables a direct comparison to be made between all PDAs, even in cases when the zeroth moment vanishes, Eq.\,\eqref{fpin0}.  The magnitudes in Fig.\,\eqref{PDAs} therefore reflect natural mass-scales associated with these PDAs.

Given that the pseudoscalar-meson leptonic decay constant discussed in connection with Eqs.\,\eqref{fpin0}, \eqref{fpigen0} is simultaneously the zeroth moment of that meson's PDA, Eq.\,\eqref{phimom}, it was always clear that the manifestation of DCSB in the PDAs of pseudoscalar meson radial excitations would be striking.  Indeed, $f_{\pi_{n\geq 1}}\equiv 0$ at $\hat m = 0$ entails that in the chiral limit the first two coefficients in any Chebyshev expansion of $\phi_{\pi_{n\geq 1}}$ are required to vanish at all resolving scales, $\zeta$.  The leading term in Eq.\,\eqref{Gfit} must then involve $C_{z_{\rm min}}^\alpha(x-\bar x)$ with $z_{\rm min}\geq 2$; and the value $z_{\rm min}=2$ largely explains the general features of the $\pi_1$ and $K_1$ PDAs in Fig.\,\ref{PDAs}.

One might guess that the PDA associated with the pion's second radial excitation should possess four zeros on $x\in(0,1)$.  That can be also achieved with a PDA of the form in Eq.\,\eqref{Gfit} using $z_{\rm min}=2$, \emph{e.g}.\ so long as that function is orthogonal to $\phi_{\pi_1}$.  Orthogonality, however, will typically produce $a^4_{\pi_2} > a^2_{\pi_2}$ in Eq.\,\eqref{Gfit}.  We have checked by direct chiral-limit computation of $m_{\pi_2}$, $\Gamma_{\pi_2}$, $f_{\pi_2}$, $\phi_{\pi_2}$; and all these expectations are confirmed.

Pursuing this line of reasoning further, we remarked in Sec.\,\ref{AdSQCD} that the ERBL evolution equations  for pseudoscalar-meson radial excitations are the same as those for ground-states.  Consequently, our expectation for the conformal-limit behaviour of the PDAs associated with radially-excited pseudoscalar mesons is
\begin{equation}
\label{pincl}
\phi_{\pi_n}^{\rm cl}(x;\zeta) = c_{\pi_n}(\zeta) \, x \bar x\, C_2^{3/2}(x-\bar x)\,,
\end{equation}
with $c_{\pi_n}(\zeta)\to 0$ as $\Lambda_{\rm QCD}/\zeta \to 0$.  Notwithstanding this, we anticipate that at $\zeta_2$ the dominant coefficient in Eq.\,\eqref{Gfit} will be associated with $a_{\pi_n}^{z=2n}$; but, since coefficients of higher Chebyshev polynomials vanish (slightly) faster under ERBL evolution, there should always be a scale, $\zeta_{\rm cl}$, such that Eq.\,\eqref{pincl} is valid $\forall \zeta > \zeta_{\rm cl}$.
Returning to Fig.\,\ref{PDAs}, in order to provide additional context for our numerical results we also plot $\phi_{\pi_1}^{\rm cl}(x)$ with $c_{\pi_1}=1.40$, chosen such that $\langle x^2\rangle$ computed using the PDA obtained therewith is the same as that listed for $\pi_1$ in Table~\ref{Table:moments}.  Notably, the prediction of the holographic model, Eq.\,\eqref{phicl}, conflicts with Eq.\,\eqref{pincl} and the associated discussion.

Important to this discussion is the feature of mutual orthogonality between the PDAs we compute for different radial excitations.  That is not a characteristic of the PDAs produced by the holographic model described in Sec.\,\ref{AdSQCD}.  Our analysis and results therefore indicate that Eq.\,\eqref{kporthogonal} is a loose assumption, not valid in general.  The general statement of orthonormality for light-front wave functions should be
\begin{equation}
\label{xkporthogonal}
\int_0^1 \! dx \! \int \frac{d^2 k_\perp}{16\pi^3} \, \Psi_{\pi_i}(x,|k_\perp|)\, \Psi_{\pi_j}(x,|k_\perp|) = \delta_{ij}\,.
\end{equation}
Thus, whilst Eq.\,\eqref{kporthogonal} is a sufficient condition, it is not necessary and hence is likely insufficient to adequately constrain a realistic wave function.

Turning now to some particular features of the $\pi_1$ and $K_1$ PDAs, the lower panel of Fig.\,\ref{PDAs} depicts $|\phi(x)|$, which may be interpreted as a probability density.  The radial-excitation curves in this panel exhibit a distinctive pattern of support: $|\phi_{\pi_1}|$ is plainly dilated with respect to $|\phi_{\pi_1}^{\rm cl}|$, with a significant amount of strength shifted to the endpoints, analogous to the pattern seen in the ground state; and $|\phi_{K_1}|$ is markedly distorted toward $x=0$ ($\bar x=1$).  These images help in understanding the top panel, which is now seen to confirm that $\phi_{\pi_1}(x)$ is dilated with respect to $\phi_{\pi_1}^{\rm cl}(x)$ and $\phi_{K_1^+}(x)$ is distorted, with its minimum located at $\bar x= 0.6$, and
\begin{subequations}
\begin{align}
\langle 2\bar x -1 \rangle_{K_1^+} &= 0.079\,f^E_{K}\,,   \\
\langle (2\bar x -1)^2 \rangle_{K_1^+} &= 0.82\,f^E_{K} \,,   \\
\langle (2\bar x -1)^2 \rangle_{\pi_1} & = 0.87\,f^E_{\pi}\,,
\end{align}
\end{subequations}
values which may be compared with $\langle (2 x -1)^2 \rangle_{\phi^{\rm cl}} = 0.2\,f^E_\pi$.  Evidently, therefore, as in the ground-state $K^+$, the $\bar s$-quark carries more of the bound-state's momentum than the lighter $u$-quark and flavour-symmetry breaking is a 20\% effect.  Moreover, here, as in so many other cases \cite{Braun:2004vf, ElBennich:2011py, Chen:2012txa, Shi:2015esa, Chen:2016snoPRD}, it is the flavour-dependence of DCSB that determines the strength of $SU(3)$-flavour breaking, not the current-quark mass-difference generated by the Higgs mechanism.

It is worth reiterating at this point that all DSE numerical results reported above were obtained using RL truncation.  Based on analyses of pseudoscalar-meson ground-states, we anticipate that the more realistic DCSB-improved truncation (see Ref.\,\cite{Binosi:2014aea} and Appendix\,A.2, Ref.\,\cite{Chang:2012cc}) will produce PDAs for the radial excitations that exhibit less dilation and less $s$-quark/$u$-quark distortion; but it won't have any material qualitative impact on our results.

In closing this section it is worth remarking that, owing to charge-conjugation invariance and the $SU(N_f)$ vector Ward-Green-Takahashi identity \cite{Maris:2000ig, Diehl:2001xe, Bhagwat:2006py, Chen:2012txa}, the leptonic decay constant is zero for any and all states on the $J^{PC} = 0^{++}$ trajectory constituted from equal-mass valence-quarks, whether ground-state, radial excitation or hybrid.  This decay constant is the vector-projection of the scalar meson's Bethe-Salpeter wave function onto the origin in configuration space, \emph{viz}.\ the zeroth moment of the scalar meson's PDA.  As a consequence, scalar meson PDAs must also exhibit interesting features.

\section{Conclusion}
\label{summary}
We computed the parton distribution amplitudes (PDAs) of the first radial excitations of the $\pi$- and $K$-mesons: $\phi_{\pi_1}(x)$ and $\phi_{K_1}(x)$, respectively, where $x$ is the valence-quark's light-front momentum fraction.  The properties and structure of ground states in these channels are strongly influenced by dynamical chiral symmetry breaking (DCSB) and this is also true of the radial excitations, in some ways more strikingly.  Like the ground-states, the excited-state PDAs are dilated with respect to the appropriate conformal-limit PDA and the distribution associated with the $K_1$-meson is skewed toward the heavier valence-(anti)quark.  In addition, however, $\phi_{\pi_1}(x)$ and $\phi_{K_1}(x)$ are not positive definite [Fig.\,\ref{PDAs}]: in each case there is a large domain of negative support, which contains the point $x=1/2$.  This feature is a remarkable, novel consequence of DCSB, which owes to the fact that DCSB requires the leptonic decay constant of all pseudoscalar-meson radial excitations to vanish in the chiral limit [Eq.\,\eqref{fpin0}].

It was possible to expose this feature herein because we used a symmetry-preserving truncation of QCD's two-valence-body bound-state problem, realised explicitly in a rainbow-ladder truncation of the Dyson-Schwinger equations (DSEs).  This framework preserves the one-loop renormalisation group behaviour of QCD, so that current-quark masses have a direct connection with the parameters in QCD's action and the dressed-quark mass-functions, $M_{s,u}(p^2)$, are independent of the renormalisation point.  Likewise, the renormalisation point can be fixed unambiguously, as in lattice-QCD.  Moreover, in working in the continuum and computing Bethe-Salpeter wave functions directly, the DSEs enable one to deliver a prediction for the pointwise behaviour of the PDAs on the full domain $x\in [0,1]$.  Capitalising on these features, we ensured that our results genuinely express the pattern of chiral symmetry breaking in QCD and, hence, the impact of DCSB is abundantly clear.

The analysis herein highlights that DCSB is expressed dramatically in the entire collection of pseudoscalar mesons constituted from light-quarks.  Consequently, so long as its impact is empirically evident in the pseudoscalar members of a given spectrum level, it is unlikely that chiral symmetry is restored in any of the hadrons that populate this level.

Our results were obtained using the simplest symmetry-preserving truncation of the two-valence-body problem.  They will not change qualitatively with the use of a more sophisticated truncation; but it is nevertheless worth documenting the quantitative changes, and delivering predictions in future that may reasonably be considered to be definitive.  Detailed study of PDAs characterising $n\geq 2$ radial excitations of pseudoscalar mesons may provide further insights, too, because, \emph{e.g}.\ the decay constants of these systems also vanish in the chiral limit, but their Bethe-Salpeter wave functions possess an additional zero with each level of excitation.  Likewise, calculation of the PDAs describing scalar mesons should prove instructive, given the symmetry constraints on their leptonic decay constants.  Additionally, a study of the PDAs characterising vector-meson radial excitations is also worthwhile.  The leptonic decay constants of these systems do not vanish, and it is thus possible that the associated PDAs will bear some similarity to those computed using quantum mechanical models.  Checking that possibility will serve, \emph{inter alia}, the valuable purpose of assisting in charting the domain of validity of such models.

\acknowledgments
We are grateful for insightful comments and suggestions from S.\,J.~Brodsky, I.\,C.~Clo\"et, G.\,F.~de~Teramond, B.~El-Bennich, C.~Mezrag, S.-X.~Qin, C.~Shi and P.\,C.~Tandy.
Work supported by:
the National Natural Science Foundation of China (contract nos.\ 11175004, 11275097, 11435001, 11475085 and 11535005);
the National Key Basic Research Program of China (contract nos.\ G2013CB834400 and 2015CB856900)
the U.S.\ Department of Energy, Office of Science, Office of Nuclear Physics, under contract no.~DE-AC02-06CH11357;
and the Chinese Ministry of Education, under the \emph{International Distinguished Professor} programme.


\end{document}